\documentclass[prc,twocolumn,showpacs,preprintnumbers]{revtex4}
\usepackage{bm}
\usepackage{amssymb}
\usepackage{amsmath}
\usepackage{graphicx}

\begin{document}
\title{Diffusion in Coulomb Crystals}
\author{J. Hughto}\email{jhughto@indiana.edu}
\author{A. S. Schneider}
\author{C. J. Horowitz}\email{horowit@indiana.edu} 
\affiliation{Department of Physics and Nuclear Theory Center,
             Indiana University, Bloomington, IN 47405}
\author{D. K. Berry}
\affiliation{University Information Technology Services,
             Indiana University, Bloomington, IN 47408}

\date{\today}
\begin{abstract}
Diffusion in coulomb crystals can be important for the structure of neutron star crusts.  We determine diffusion constants $D$ from molecular dynamics simulations.  We find that  $D$ for coulomb crystals with relatively soft-core $1/r$ interactions may be larger than $D$ for Lennard-Jones or other solids with harder-core interactions.  Diffusion, for simulations of nearly perfect body-centered-cubic lattices, involves the exchange of ions in ring-like configurations.  Here ions ``hop" in unison without the formation of long lived vacancies.  Diffusion, for imperfect crystals, involves the motion of defects.  Finally, we find that diffusion, for an amorphous system rapidly quenched from coulomb parameter $\Gamma=175$ to coulomb parameters up to $\Gamma=1750$, is fast enough so that the system starts to crystalize during long simulation runs.  These results strongly suggest that coulomb solids in cold white dwarf stars, and the crust of neutron stars, will be crystalline and not amorphous.
\end{abstract}
\smallskip


\maketitle

\section{Introduction}
\label{sec.introduction}
Diffusion in coulomb plasma {\it liquids}  has been well studied \cite{Hanson75} and is important for sedimentation of impurities in white dwarf (WD) \cite{Ne_lars, Ne_lars2, neon_diffusion} and neutron stars (NS) \cite{NS_diffusion, peng}.  Here ions, with a larger than average mass to charge ratio, sink in a strong gravitational field.  This releases gravitational energy that can delay the cooling of metal rich WD \cite{WD_cooling}.  However, we are not aware of  numerical results for diffusion constants of coulomb {\it crystals} under Astrophysical conditions.  Often the diffusion constant is simply assumed to be zero.  This diffusion could be important for sedimentation in solid WD interiors, over long time scales, and for the structure of NS crusts.

Solid diffusion can depend dramatically on the form of the interaction between particles and may be very slow for hard-core systems.  For example, the binary Lennard Jones (LJ) system with a hard-core $\propto r^{-12}$ interaction forms a glass because of very slow diffusion \cite{LJglass}.  In contrast, the coulomb plasma with a soft $1/r$ core should have much faster diffusion.  Therefore the Coulomb crystal may provide an important model system where diffusion is fast enough to be more easily studied by molecular dynamics (MD) simulations.

In the laboratory, one can observe diffusion in complex (or dusty) plasma crystals.  Complex plasmas (CP) are low temperature plasmas containing charged microparticles, for a review see Fortov et al. \cite{fortov}.  Often the microparticles are micron sized spheres that acquire large electric charges and the strong coulomb interactions between microparticles can lead to crystallization.  Indeed plasma crystals were first observed in the laboratory in 1994 \cite{dusty_plasma}.   Complex plasmas typically differ from White Dwarf interiors and Neutron Star crusts in a number of ways.  First the microparticles feel additional fluctuating and friction forces because of interactions with the background gas.  Note that in stars, electron-ion interactions are small because of the large electron degeneracy.  Second, the Debye screening length $\lambda$, see Eq. \ref{v(r)} below, is often smaller in the CP than in a star (when measured in units of the lattice spacing).  This changes the lattice type from body-centered-cubic (bcc) as expected in stars, to face-centered-cubic (fcc) or other types in a CP.  Finally in a CP there is an overall confining potential, and because of gravitational gradients it is often easier to study two-dimensional CP crystals.  

In two dimensions, one can have liquid, crystalline, and semi-crystalline states.  Anomalous diffusion in semi-crystalline CP states has been observed at intermediate times  \cite{2dprl,NJP}.  In anomalous diffusion the square of the displacement does not grow linearly with time.  Langevin-dynamics simulations \cite{2dtheory} find that microparticle-background gas interactions are important for this diffusion. 

The melting of colloidal crystal films has recently been studied \cite{peng2010}.  Thick films ($>$ 4 layers) were observed to melt at grain boundaries, while films 2 to 4 layers thick melted from both grain boundaries and from within crystalline domains.  We study diffusion at grain boundaries in Sec. \ref{subsec.frozen}.

Three-dimensional CP crystals have been formed onboard the International Space Station under microgravity conditions.  Details of the experiment are presented in ref. \cite{ISS1}.  The structural properties of the crystal were analyzed with bond angle metrics $q_4$ and $q_6$, see Section \ref{subsec.amorphous}.  Microparticles were found in regions with fcc and hexagonal-close packing (hcp) order \cite{ISS2,ISS3}, in agreement with MD simulations \cite{ISS2}.   Khrapak et al. \cite{khrapak} studied freezing and melting of these CP crystals and found diffusion to be relatively fast so that the system remained in equilibrium.  Melting criteria for CP systems were presented by Klumov \cite{klumov0}. 

We now focus on simple plasmas in three dimensions.  The diffusion mechanism is interesting.  Astrophysical systems are under great pressure that suppresses the formation of vacancies.  Therefore diffusion, in a nearly perfect crystal, should involve the exchange of neighboring ions.  These exchanges, while common in some quantum systems, may be less common in classical systems.  More complicated coulomb solids can involve a variety of dislocations, grain boundaries, and other imperfections.  Diffusion in these systems probably involves motion of the imperfections, since this may be faster than particle exchanges.  Determining the diffusion constant for a system may help characterize the kinds and numbers of imperfections.  Note that the coulomb plasma has especially simple interactions.  Therefore, it may be a very useful model system to study diffusion in the presence of complex imperfections.

We emphasize that the coulomb plasma has no hard core interaction between ions, but only a relatively weak $1/r$ repulsion.  Therefore, it may be possible for ions to come relatively close to one another, if necessary for the motion of defects.  This may be different from conventional condensed matter with hard cores.  For example, MD simulations of defect motion in Magnesium focused on paths that involved only very small displacements of Magnesium atoms \cite{magnesium}.  Imperfections may move much faster in a coulomb plasma.

The motion of imperfections is important for equilibration.  For example, a coulomb liquid may freeze  into an imperfect crystal state involving an excess of defects.  There has been some work on nucleation in coulomb plasmas, see for example \cite{nucleation}.  However present MD simulations of nucleation may have limitations from important finite size effects \cite{nucleationMD}.  In this paper, we also study diffusion in amorphous systems to see if it is fast enough to allow crystallization.

We focus on one component plasmas (OCP).  We plan to study diffusion in multicomponent plasmas (MCP) in the future.  As we discuss below, this may help address an important unsolved problem, the structure of MCP crystals.  This is important for the thermal and electrical conductivity of NS crust  \cite{thermal_conductivity}.  Indeed X-ray observations of rapid NS crust cooling, after extended periods of accretion, strongly favor the formation of a crystalline rather than amorphous crust and may set limits on impurities \cite{NScrustcooling,crustcooling2,crustcooling3,crustcooling4}.  In addition, pycnonuclear reactions, which are driven by quantum zero point motion at high densities, are exponentially sensitive to the structure of MCP crystals and the spatial locations of reactants \cite{pycnonuclear}.  These reactions may provide an important heat source in the crust of accreting NS \cite{pycnofusion}.  Finally the distribution of dislocations, grain boundaries, impurities, and other imperfections are important for mechanical properties of NS crust such as its breaking strain \cite{breakingstrain1,breakingstrain2}.  The breaking strain helps determine the maximum sized mountains that are possible on a NS, which are important for gravitational wave radiation \cite{breakingstrain1,GWlowmass}.  The breaking strain also determines the maximum sized ``star quake'' that is possible.  Sudden changes, or glitches, in the rotational period of pulsars \cite{glitches} may involve crust breaking that could trigger the motion of superfluid vortices.  In addition Magnetar giant flares, extremely intense gamma ray flares from very strongly magnetized NS \cite{giantflares}, may involve the catastrophic breaking of the crust because of very large magnetic stresses \cite{magnetars}.             

In previous work we determined liquid-solid phase equilibrium for a MCP system involving many ion species \cite{phasesep}, see also \cite{WDsep}.  We performed a large scale MD simulation where both liquid and solid phases were present.  The solid phase in this simulation may have had a number of imperfections.  A knowledge of diffusion constants $D$ may help determine the simulation time necessary for these imperfections to come into equilibrium.

There have been previous calculations of $D$ for coulomb liquids, starting with the MD simulations of Hansen et al. for the one component plasma (OCP) \cite{Hanson75}.  The one component plasma consists of ions, with pure coulomb interactions, and an inert neutralizing background charge density.  Diffusion in the OCP in a strong magnetic field was considered by Bernu \cite{Bernu}.  Hansen et al. have also calculated diffusion for binary mixtures \cite{Hanson85}.           

Diffusion for a Yukawa fluid has been simulated by Robbins et al. \cite{Robbins} and Ohta et al. \cite{Ohta}.  In a Yukawa fluid ions interact via a screened coulomb potential $v_{ij}(r)$,
\begin{equation}
v_{ij}(r)=\frac{Z_iZ_j e^2}{r} {\rm e}^{-r/\lambda},
\label{v(r)}
\end{equation}
for two ions with charges $Z_i$ and $Z_j$, that are separated by a distance $r$. The OCP is equivalent to a Yukawa fluid, where all of the ions have the same charge and the screening length $\lambda$ is very large.

The motion of ions in a WD or NS is largely classical because of their large mass.  However at great densities, there could be quantum corrections that might increase $D$.  These have been estimated for a liquid by Daligault and Murillo \cite{Daligault}, and found to be very small.    

In this paper, we present classical MD simulations of one component crystals with Yukawa interactions in order to determine diffusion coefficients $D$.   In Section \ref{sec.formalism} we describe our MD formalism and present results for diffusion coefficients in Section \ref{sec.results}.  We conclude in Section \ref{sec.conclusion}.

\section{Formalism}
\label{sec.formalism}

We describe our MD simulation formalism.  This is similar to what we used earlier to calculate $D$ for liquid mixtures of carbon, oxygen, and neon \cite{neon_diffusion}.  We consider a one component system of oxygen ions where the ions are assumed to interact via screened Yukawa interactions, see Eq. \ref{v(r)}.  The Thomas Fermi screening length $\lambda$, for cold relativistic electrons, is 
\begin{equation}
\lambda^{-1}=2\alpha^{1/2}k_F/\pi^{1/2} 
\end{equation}
where the electron Fermi momentum $k_F$ is $k_F=(3\pi^2n_e)^{1/3}$ and $\alpha$ is the fine structure constant.  The electron density $n_e$ is equal to the ion charge density, $n_e=Z n$, where $n$ is the ion density and $Z$ is the ion charge.  Our simulations are classical and we have neglected the electron mass (extreme relativistic limit).   This is to be consistent with our previous work on neutron stars.  However, the electron mass is important at lower densities in WD and this will decrease $\lambda$.  For relativistic electrons, the ratio of $\lambda$ to the ion sphere radius $a$,
\begin{equation}
a=\Bigl(\frac{3}{4\pi n}\Bigr)^{1/3},
\end{equation}
depends only on the charge $Z$ and is independent of density.    For nonrelativistic electrons $\lambda/a$ can be somewhat smaller.  In Section \ref{sec.results}, we perform simulations for two values of $\lambda/a$.

The simulations can be characterized by a coulomb parameter $\Gamma$,
\begin{equation}
\Gamma= \frac{Z^2 e^2}{a T}\, .
\label{gammamix}
\end{equation} 
Here $T$ is the temperature.  The system freezes near $\Gamma=175$ \cite{pot1}.   Note that this value of $\Gamma$ may depend slightly on $\lambda$ \cite{hamaguchi,pot1}.
 
Time can be measured in our simulations in units of one over the plasma frequency $\omega_p$.  Long wavelength fluctuations in the charge density can undergo oscillations at the plasma frequency.  This depends on the ion charge $Z$ and mass $M$,  
\begin{equation}
\omega_p=\Bigl[\frac{4\pi e^2 Z^2 n}{ M }\Bigr]^{1/2}.
\label{omega}
\end{equation}

The diffusion constant $D$ can be calculated from the velocity autocorrelation function $Z(t)$,
\begin{equation}
Z(t)=\frac{\langle {\bf v}_j(t_0+t)\cdot {\bf v}_j(t_0) \rangle} {\langle {\bf v}_j(t_0)\cdot {\bf v}_j(t_0) \rangle}
\label{Z(t)}
\end{equation}
where the average is over all ions $j$ and over initial times $t_0$.  The velocity of the $j$th ion at time $t$ is ${\bf v}_j(t)$.  The diffusion constant is the time integral of $Z(t)$,
\begin{equation}
D=\frac{T}{M}\int_0^{t_{max}}dt Z(t).
\label{DZ(t)}
\end{equation} 
This Eq. works well to calculate $D$ for liquids.  However for crystals, $D$ is smaller and the integral in Eq. \ref{DZ(t)} involves sensitive cancelations between regions where $Z(t)$ is positive and negative. This makes Eq. \ref{DZ(t)} very difficult to use.

Instead $D$ can also be calculated from
\begin{equation}
D(t)=\frac{\langle |{\bf r}_j(t+t_0)-{\bf r}_j(t_0)|^2\rangle}{6t}
\label{D(t)}
\end{equation}
where the diffusion constant $D$ is the large time limit of $D(t)$,
\begin{equation}
D={\rm lim}_{t\rightarrow\infty} D(t).
\label{D}
\end{equation}
Here ${\bf r}_j(t)$ is the position of the $j$th ion at time $t$ and the average in Eq. \ref{D(t)} is over all ions $j$ and initial times $t_0$.  In principle, Eqs. \ref{D(t)},\ref{D} will have errors at large times $t$ from the effects of periodic boundary conditions as $|{\bf r}_j(t+t_0)-{\bf r}_j(t_0)|$ becomes comparable to the size of the simulation volume.  However diffusion is relatively slow so this is often not a problem until very large $t$. 

Note that $D(t)$ can differ significantly from $D$ for small $t$.  For example, an ion undergoing thermal oscillations about an equilibrium lattice site will have ${\bf r}_j(t)-{\bf r}_j(0)$ nonzero even though the ion remains near its original lattice site and there is no net contribution to diffusion.  Therefore we define an alternative quantity $D'(t)$ that has no contribution from ions that remain near their original lattice site,
\begin{equation}
D'(t)=\frac{\langle \Theta[|{\bf r}_j(t')-{\bf r}_j(t_0)|-R_c] |{\bf r}_j(t')-{\bf r}_j(t_0)|^2\rangle}{6 t},
\label{D'}
\end{equation}
with $t'=t+t_0$.  The cutoff radius $R_c$ is of order the lattice spacing, and will be discussed in Section \ref{sec.results}.  In the limit of very large times all ions move significantly so that $D'(t)\rightarrow D(t)$ as $t\rightarrow\infty$.  We observe that $D'(t)$ is approximately independent of $t$, even for moderate $t$, so that
\begin{equation}
D\approx D'(t)\, .
\end{equation}
We use this equation, at finite $t$, to calculate $D$ in Section \ref{sec.results}.

The initial conditions are very important for determining $D$ because the system may contain different distributions of defects and these distributions may take a very long time to equilibrate.  We consider three classes of initial conditions.  The first class we call bcc and starts the ions with positions on a perfect body centered cubic (bcc) lattice and random thermal velocities.  This may underestimate the role of defects if there is not enough simulation time for thermal excitations to introduce an equilibrium distribution of defects.  The second class of initial conditions we call imperfect crystal and starts the system from a liquid configuration that is cooled by rescaling the velocities until the system freezes.  This may over estimate the role of defects if the system freezes into a very imperfect state with more defects than would be present in thermal equilibrium.  Note that imperfect crystal initial conditions may contain two or more micro-crystals with different orientations.  Finally, we consider amorphous initial conditions where a liquid configuration is rapidly quenched to a much lower temperature.

We evolve the system in time using the simple velocity Verlet algorithm \cite{verlet}.  We approximately maintain the system at constant temperature by simply rescaling the velocities every ten time steps.  In Section \ref{sec.results} we present results for $D$.  

\section{Results}
\label{sec.results}
\begin{figure}[ht]
\begin{center}
\includegraphics[width=3.5in,angle=0,clip=true] {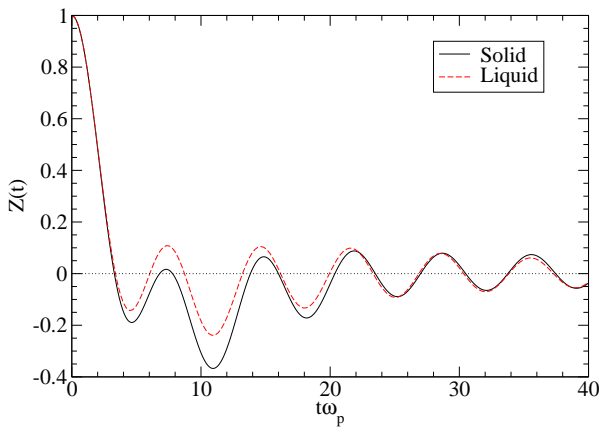}
\caption{(Color on line) Velocity autocorrelation function $Z(t)$ versus time $t$ in units of one over the plasma frequency $\omega_p$ for $N=8192$ ions at $\Gamma=176$ for both a liquid configuration (red dashed line) and a solid configuration (black solid line).}
\label{Fig1}
\end{center}
\end{figure}

We now present results for our MD simulations.  We begin with a few results insensitive to initial conditions and then we discuss simulations with perfect lattice initial conditions in Section \ref{subsec.bcc}, imperfect crystal initial conditions in Section \ref{subsec.frozen}, and amorphous initial conditions in Section \ref{subsec.amorphous}.   We start with the velocity autocorrelation function $Z(t)$, see Eq. \ref{Z(t)}, that is shown in Fig. \ref{Fig1}.  There are only subtle differences in $Z(t)$ between liquid and solid phases.  For the solid $Z(t)$  is slightly more negative for $4<t\omega_p<14$.  However this slight difference leads to a much smaller $D$ from the integral in Eq. \ref{DZ(t)}. 

Next, Fig. \ref{Fig2} shows histograms of displacements $|{\bf r}_j(t+t_0)-{\bf r}_j(t_0)|$ after a time $t=21000/\omega_p$.  These are computed by simply counting the number of ions that have moved a given distance.  Figure \ref{Fig2} shows a large peak at small distances that corresponds to ions which remain near their original lattice site.  The width of this peak corresponds to thermal oscillations.  The amplitude of these oscillations are relatively large because the system is warm and near the melting temperature.  Figure \ref{Fig2} also shows smaller peaks at larger distances that correspond to ions which have ``hopped'' one lattice site, two lattice sites, etc.  Diffusion is seen to be larger for a system that started from imperfect crystal initial condition compared to a system that started from a perfect bcc lattice initial condition.  We start by presenting additional results for perfect body centered cubic lattice initial conditions and then we will present results for imperfect crystal and amorphous initial conditions.

\subsection{Body centered cubic lattice initial conditions}
\label{subsec.bcc}

How do the ions actually move (hop) from one lattice site to the next?  This is nontrivial because the system is under high pressure and vacancy formation is suppressed.  Thus there are very few empty sites for the ions to hop into.  Instead the ions can exchange with their neighbors.  In Figure \ref{Fig3} we show the final configuration for a small 3456 ion system that was prepared from perfect bcc lattice initial conditions.  Most ions remain near their original lattice site and are shown as small brown dots.  These ions show oscillations about the lattice sites.  However for this example, there were 24 ions that moved more than $1.34a$ during the finial simulation time of $t=236/\omega_p$.  These ions are shown as larger black disks and are seen to be in a ring configuration where ions ``hop'' to lattice sites vacated by other hopping ions.    

\begin{figure}[ht]
\begin{center}

\includegraphics[width=3.5in,angle=0,clip=true] {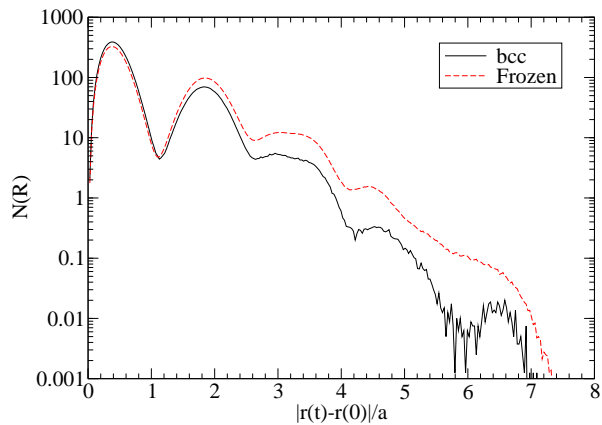}

\caption{(Color on line) Histogram of displacements $|{\bf r}_j(t+t_0)-{\bf r}_j(t_0)|$ in units of the ion sphere radius $a$ after a time $t=21000/\omega_p$.  The simulations use $N=8192$ ions and are at $\Gamma=176$.  The black solid line is the average of 800 configurations (initial times $t_0$) for a system that started from a perfect body centered cubic (bcc) lattice initial configuration while the dashed red line is the average of 8000 configurations for a system that started from an imperfect crystal initial configuration. }
\label{Fig2}
\end{center}
\end{figure}

\begin{figure}[ht]
\begin{center}
\vskip.1in
\includegraphics[width=3.25in,angle=0,clip=true] {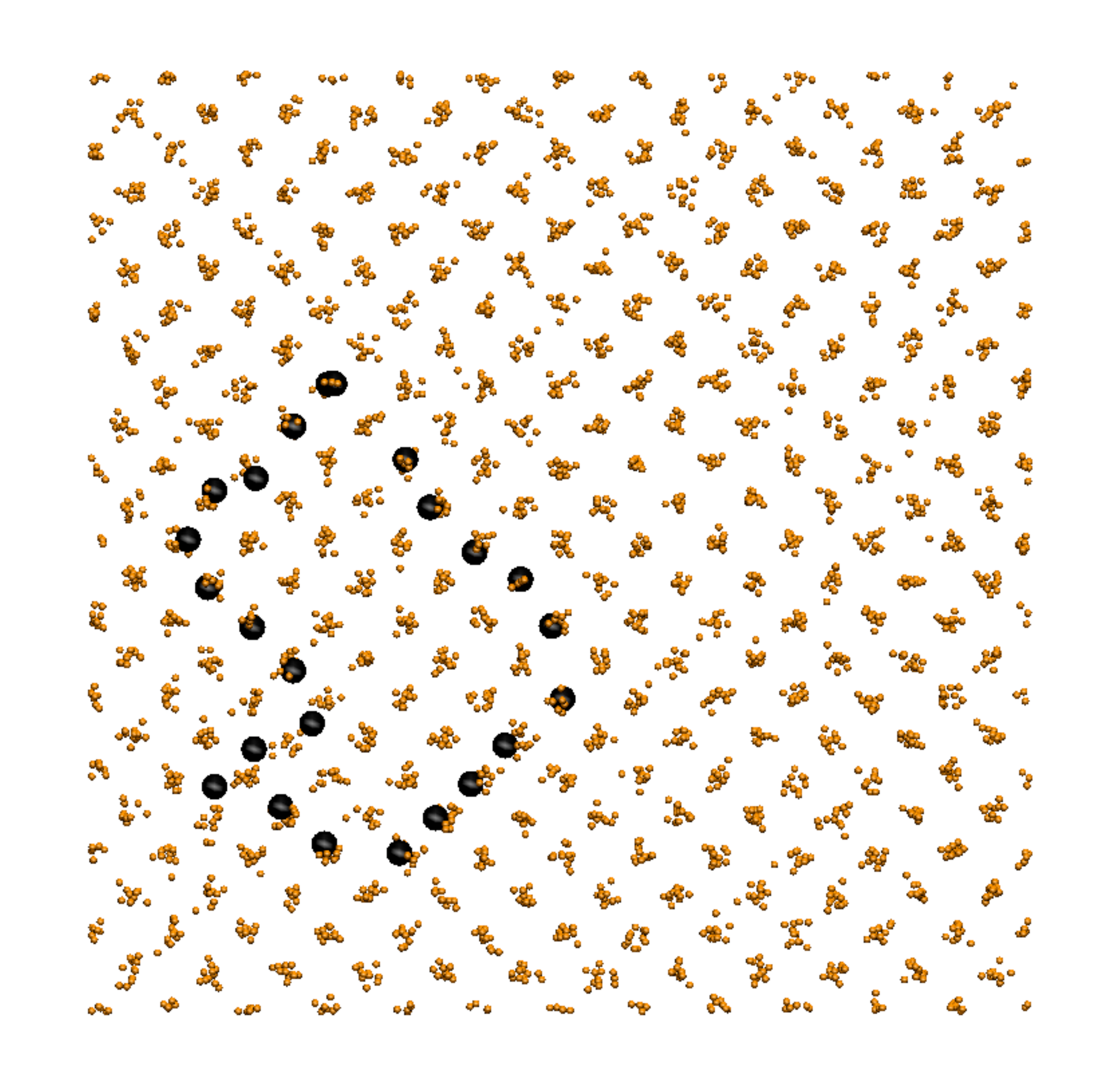}
\caption{(Color on line) Sample configuration of 3456 ions at $\Gamma=175$.  Ions that have moved less than $1.34a$ in a time $t=236/\omega_p$ are small brown dots.  Ions that have moved more than $1.34a$ are shown as larger black disks and are seen to be in a ring configuration where ions ``hop'' to lattice sites vacated by other hopping ions.  This system started from a perfect bcc lattice.  Figure plotted with VMD \cite{VMD}. }
\label{Fig3}
\end{center}
\end{figure}

We now present results for the diffusion constant $D$ using Eq. \ref{D'} with a cutoff parameter $R_c$ chosen as the location of the minimum in the histograms in Fig. \ref{Fig2} at,
\begin{equation}
R_c=1.07a\, .
\end{equation}
Small changes in this value only lead to slight changes in $D$.  To minimize finite size effects we also introduce a cutoff range $R_{\rm cut}$ in the Yukawa interaction so that Eq. \ref{v(r)} becomes
\begin{equation}
v_{ij}=Z_iZ_j e^2\Bigl[\frac{{\rm e}^{-r/\lambda}}{r}-\frac{{\rm e}^{-R_{\rm cut}/\lambda}}{R_{\rm cut}}\Bigr]\Theta(R_{\rm cut} -r)
\end{equation}
and the potential is zero for $r>R_{\rm cut}$.

 \begin{table}
\caption{Diffusion constant $D$ for MD simulations starting from perfect body centered cubic lattice initial conditions at $\Gamma=175$.  Here $D$ is in units of $\omega_p a^2$ with $\omega_p$ the plasma frequency and $a$ the ion sphere radius, $N$ is the number of ions, $\lambda$ the screening length, $R_{\rm cut}$ the cutoff radius in the interaction, $\Delta t$ the MD time step, $t$ the elapsed time, and $N_{\rm conf}$ the number of configurations used to average over the inital time $t_0$.}
\begin{tabular}{lllllll}
 $N$ & $\lambda/a\ \ $ & $R_{\rm cut}/\lambda$ & $\Delta t\omega_p\ \ $ & $t\omega_p$ & $N_{\rm config}$ & $D/\omega_p a^2$\\
 \hline
3456 & 1.82 & $\infty$ & 0.047 & 35000 & 700 & $6.2\times 10^{-7}$\\
3456 & 2.70 & $\infty$ & 0.047 &47200 & 200 & 0\\

8192 & 1.82 & 8.91 & 0.12 & 170000 & 800 & $7.7\times 10^{-6}$\\
8192 & 2.70 & $\infty$ & 0.12 & 106000 & 800 & $3.4\times 10^{-7}$\\

27648 & 1.82 & 8.91 & 0.12 & 59000 & 500 & $1.0\times 10^{-5}$ \\
27648 & 1.82 & $\infty$ & 0.12 & 59000 & 1200 & $1.0\times 10^{-5}$ \\
27648 & 2.70  & $\infty$ & 0.06 & 23600 & 89 & $4.9\times 10^{-6}$ \\
27648 & 2.70 & $\infty$ & 0.06 & 668500 & 396 & $4.5\times 10^{-6}$ \\
27648 & 2.70 & $\infty$ & 0.06 & 23600 & 300 & $5.1\times 10^{-6}$ \\
27648 & 2.70 & $\infty$ & 0.12 & 59000 & 500 & $4.2\times 10^{-6}$ \\
27648 & 2.70 & $\infty$ & 0.12 & 23600 & 800 & $4.5\times 10^{-6}$ \\
27648\footnote{This is a continuation of the run described in the line above.  However, it is at constant energy instead of being at (approximately) constant temperature.}& 2.70 & $\infty$ & 0.12 & 23600 & 800 & $4.8\times 10^{-6}$ \\
27648 & 2.70 & $\infty$ & 0.24 & 23600 & 300 & $4.9\times 10^{-6}$ \\

93312 & 1.82 & 8.91 & 0.12 & 59000 & 350 & $1.0\times 10^{-5}$\\
93312 & 2.70 & $\infty$ & 0.12 & 59000 & 350 & $5.6\times 10^{-6}$\\
\end{tabular} 
\label{tableone}
\end{table}

We first consider bcc lattice initial conditions.  Table \ref{tableone} presents results for $D$ for different values of $\lambda$, molecular dynamics time step $\Delta t$, number of ions $N$, and elapsed time $t$ used in Eq. \ref{D'}.  We express $D$ in units of $\omega_p a^2$.  We find that $D$ increases with decreasing $\lambda$.  For a large value of $\lambda=2.70a$ there are large finite size effects and $D$ increases with increasing $N$.  However the increase in $D$ in going from $N=27648$ to the largest system size 93312 is small.  

Finite size effects are smaller for the smaller $\lambda=1.82a$ value.  Now there is good agreement for $N=27648$ and 93312 and $D$ is only slightly smaller for $N=8192$.  We do not find strong sensitivity to $\Delta t$ or $t$.  Furthermore, imposing a cutoff on the interaction at large distances $R_{\rm cut}=8.91\lambda$ has only a very small effect on $D$.  For $\lambda=1.82a$ and large systems, we find $D/\omega_pa^2=1.0\times 10^{-5}$.  As we discuss below, this value, for the solid near the melting point $\Gamma=175$, is about 200 times smaller than $D$ for the liquid phase at the same $\Gamma$.

\begin{table}
\caption{Diffusion constant $D$ versus $\Gamma$ for MD simulations using $N=27648$ ions and starting from perfect body centered cubic lattice initial conditions.  The screening length is $\lambda$, $R_{\rm cut}$ is the cutoff radius for the interaction, the MD time step is $\Delta t\omega_p=0.12$ and $t=59000/\omega_p$.}
\begin{tabular}{llll}
 $\Gamma$ & $\lambda/a\ \ $ & $R_{\rm cut}/\lambda$  & $D/\omega_p a^2$\\
 \hline
165 & 1.82 & 8.91 & $3.3\times 10^{-5}$\\
175 & 1.82 & 8.91  & $1.05\times 10^{-5}$\\
185 &  1.82 &  8.91 & $3.3\times 10^{-6}$\\
200 &1.82   & 8.91  & $3.9\times 10^{-7}$\\

165 & 2.70 & $\infty$ & $1.3\times 10^{-5}$\\
175 & 2.70 & $\infty$ & $4.1\times 10^{-6}$\\
185 & 2.70 & $\infty$  & $1.6\times 10^{-6}$\\
200 & 2.70  &$\infty$  & $1.5\times 10^{-7}$\\
\end{tabular} 
\label{tabletwo}
\end{table}

Diffusion in the solid may involve an energy barrier $\Delta E$ since it may be necessary for an ion to pass close to its neighbors.  This would lead to a temperature dependance $D\propto {\rm Exp}(-\Delta E/T)={\rm Exp}(-d\, \Gamma)$ with $d$ a constant.  In Table \ref{tabletwo} we present results for $D$ as a function of $\Gamma$.  For $\Gamma\le 185$, Table \ref{tabletwo} results are approximately 
\begin{equation}
\frac{D}{\omega_pa^2}\approx 6100\, {\rm Exp}(-0.115\, \Gamma)
\end{equation}
for $\lambda=1.82 a$ and  
\begin{equation}
\frac{D}{\omega_pa^2}\approx 1400\, {\rm Exp}(-0.112\, \Gamma)
\end{equation}
for $\lambda=2.70a$.  Note that $\Delta E$ (or $d\approx 0.11$) appears to be almost independent of $\lambda$.  This would follow if $\Delta E$ is dominated by particle interactions at short distances. 
\begin{figure}[ht]
\begin{center}
\includegraphics[width=3.5in,angle=0,clip=true] {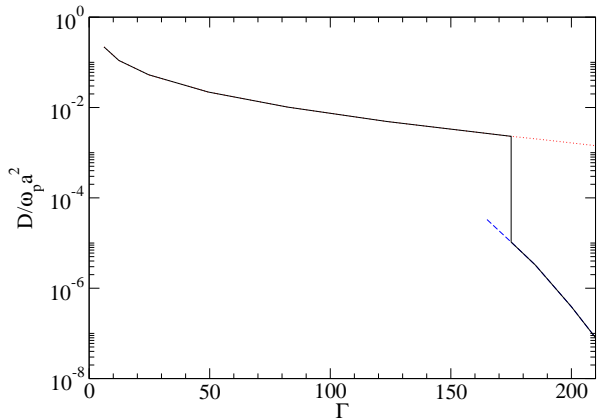}
\caption{(Color on line) Diffusion constant $D$ versus $\Gamma$ for both the liquid and solid phases.  Liquid results are from ref. \cite{neon_diffusion} while the solid results are for $\lambda=1.82a$ and assume perfect bcc lattice initial conditions.  The dotted red line shows metastable (super cooled) liquid results while the blue dashed line shows metastable (super heated) solid results.  The system is assumed to melt at $\Gamma=175$. }
\label{Fig4}
\end{center}
\end{figure}

In Fig. \ref{Fig4} we plot $D$ as a function of $\Gamma$ for both the liquid and solid phases.  We see that $D$ drops by a large factor as the system crystalizes and that $D$ decreases much more rapidly, with increasing $\Gamma$, in the solid phase compared to the behavior of $D$ in the liquid phase.

Most of our simulations are at (approximately) constant temperature where velocities are rescaled every ten time steps to keep the kinetic energy fixed.  To test the sensitivity of our results to this procedure, we have also performed a few runs at constant energy, instead of at constant temperature, see for example Table \ref{tableone}.  We find that $D$ is unchanged within statistics.

\subsection{Imperfect crystal initial conditions}
\label{subsec.frozen}

We now consider imperfect crystal initial conditions.  We prepare a liquid initial condition by starting the ions off at random positions, with a Maxwell velocity distribution, and evolving the system at a series of increasing $\Gamma$ values.  The system is observed to equilibrate in a liquid phase.  However as $\Gamma$ is increased further the system is observed to supercool for $\Gamma>175$ and then eventually freeze.  However, often the system freezes into an imperfect crystal with many defects.  For example, there can be two micro-crystals of different orientations.  Once the system has frozen, $\Gamma$ is decreased back to $\Gamma=175$ and the system is evolved for a long time at this $\Gamma$ value and the diffusion constant is calculated from Eq. \ref{D'}.  

\begin{figure}[ht]
\begin{center}
\vskip.1in
\includegraphics[width=3.25in,angle=0,clip=true] {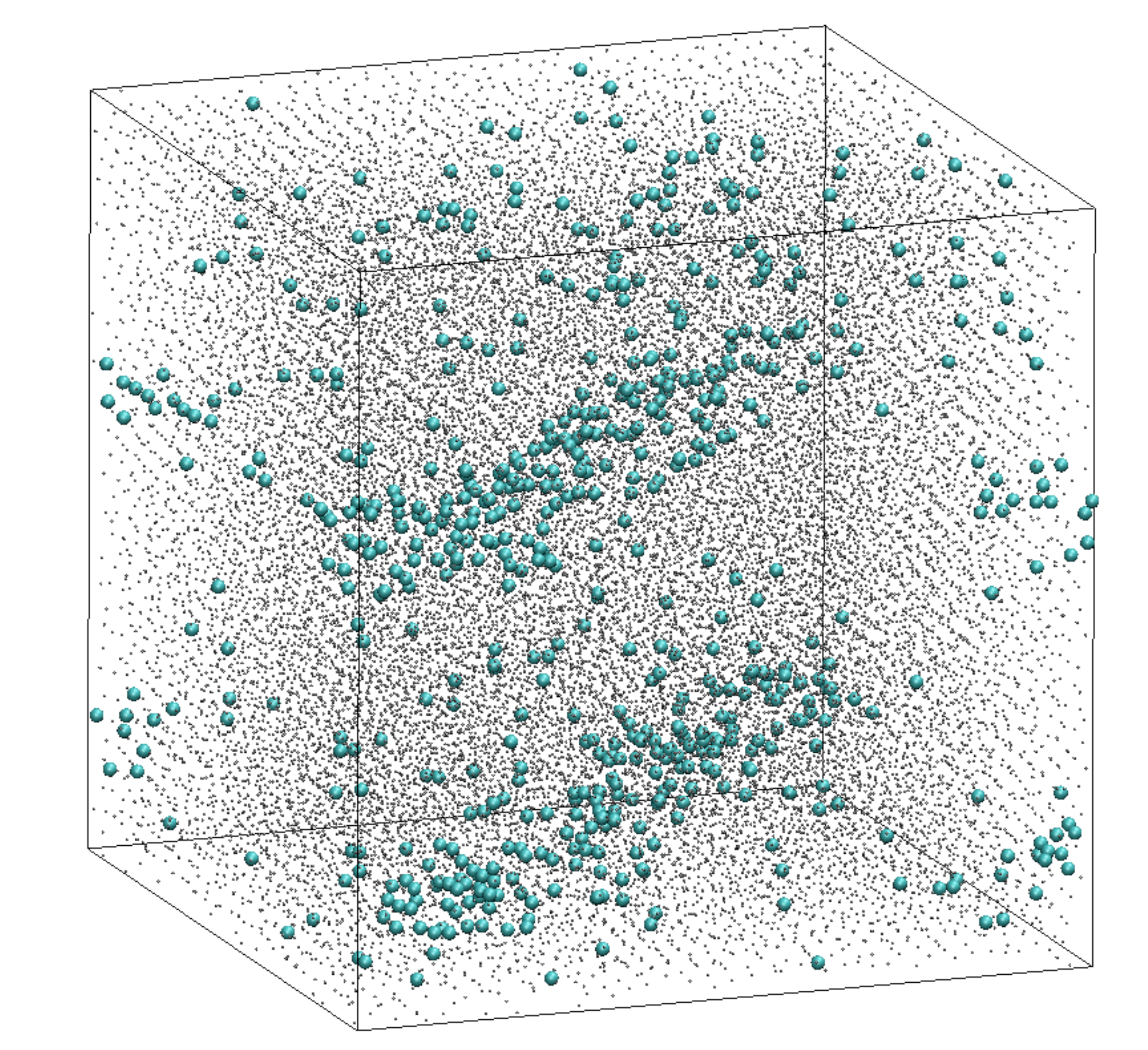}
\caption{(Color on line) Configuration of 27648 ions starting from imperfect crystal initial conditions.  Ions that move only a small distance are small gray points.  Ions that have moved over three lattice spacings, during the simulation time of $t=59000/\omega_p$, are shown as large blue spheres.  These are seen to be clustered at the grain boundaries.  The initial conditions included two micro-crystals of different orientation.  Figure plotted using VMD \cite{VMD}. }
\label{Fig5}
\end{center}
\end{figure}

Figure \ref{Fig5} shows a configuration of 27648 ions with imperfect crystal initial conditions.  Here ions, that have moved less than three lattice spacings during the simulation, are plotted as small gray points while ions, that have moved more than three lattice spacings, are plotted as large blue spheres.   The system froze into two micro-crystals of different orientation and the diffusing ions are seen to be clustered near the grain boundaries.  This suggests that diffusion in imperfect crystals may be dominated by motion of the defects rather than by hopping of ion chains, such as that shown in Fig. \ref{Fig3}.

 \begin{table}
\caption{Diffusion constant $D$ for MD simulations starting from imperfect crystal initial conditions at $\Gamma=175$.  Here $D$ is in units of $\omega_p a^2$ with $\omega_p$ the plasma frequency and $a$ the ion sphere radius, $N$ is the number of ions, $\lambda$ the screening length, $R_{\rm cut}$ the cutoff radius in the interaction, $\Delta t$ the MD time step, $t$ the elapsed time, and $N_{\rm config}$ the number of configurations used to average over the initial time $t_0$.}
\begin{tabular}{lllllll}
 $N$ & $\lambda/a\ \ $ & $R_{\rm cut}/\lambda$ & $\Delta t\omega_p\ \ $ & $t\omega_p$ & $N_{\rm config}$ & $D/\omega_p a^2$\\
 \hline
3456 & 2.70 & $\infty$ & 0.047 & 4700 & 380 & $4.9\times 10^{-5}$ \\

8192 & 2.70 & $\infty$ & 0.12 & 9600 & 8000 & $1.4\times 10^{-5}$\\

27648 & 2.70 & $\infty$ & 0.12 & 59000 & 400 & $8.9\times 10^{-6}$ \\
27648\footnote{This is a continuation of the run described in the line above.  However it is at constant energy instead of being at (approximately) constant temperature.} & 2.70 & $\infty$ & 0.12 & 59000 & 400 & $8.8\times 10^{-6}$\\
27648 & 2.70  & $\infty$ & 0.12 & 59000 & 800 & $1.9\times 10^{-5}$ \\
27648 & 1.82 & 8.91 & 0.12 & 59000 & 500 & $2.3\times 10^{-5}$ \\
27648 & 1.82 & $\infty$ & 0.06 & 472000 & 401 & $1.1\times 10^{-5}$ \\

\end{tabular} 
\label{tablethree}
\end{table}

In Table \ref{tablethree} we present results for $D$ for imperfect crystal initial conditions.  Note that lines 3 and 4 in Table \ref{tablethree} and lines 5 and 6 correspond to independently prepared initial conditions.  There is some variation in results for different simulations.  This may reflect differences in the number and kind of defects present in the initial conditions.  We see that $D$ for imperfect crystal initial conditions is two to four times larger than $D$ for perfect bcc lattice initial conditions.  We also see that $D$ may be less sensitive to the screening length for imperfect crystal initial conditions.

It is possible that $D$ will evolve slowly with simulation time $t_0$ for these imperfect crystal simulations.  Note that we do not find rapid variation of $D$ with $t_0$.  However, we have not attempted to determine how $D$ might evolve over long times by continuing an imperfect crystal simulation for very long times.     

The final simulation listed in Table \ref{tablethree} was prepared by very slowly cooling a liquid configuration that started at $\Gamma=175$, at a rate of $d\Gamma/dt=2.1\times 10^{-4}\omega_p$, until the configuration froze at $\Gamma=283$.  The resulting solid configuration was then heated back up to $\Gamma=175$.  Finally, the system was evolved at $\Gamma=175$ for a time $t_0=59000/\omega_p$ before taking $D$ data.  This system was observed to be a nearly perfect bcc lattice, and the value for $D$ in Table \ref{tablethree} agrees with our results for nearly perfect bcc lattices in Section \ref{subsec.bcc}.  This strongly suggests that white dwarf and neutron star plasmas will freeze into nearly perfect bcc crystals, since any astrophysical cooling time scale is likely very much longer than this MD cooling time scale.  This is also consistent with our results in Section \ref{subsec.amorphous} for amorphous systems, see below.       

\begin{table}
\caption{Diffusion constant $D$ versus $\Gamma$ for MD simulations using $N=27648$ ions and starting from a single imperfect crystal initial condition.  The screening length is $\lambda$, $R_{\rm cut}$ is the cutoff radius for the interaction, the MD time step is $\Delta t\omega_p=0.12$ and $t=59000\omega_p$.}
\begin{tabular}{llll}
 $\Gamma$ & $\lambda/a\ \ $ & $R_{\rm cut}/\lambda$  & $D/\omega_p a^2$\\
 \hline
175 & 1.82 & 8.91  & $2.3\times 10^{-5}$\\
185 &  1.82 &  8.91 & $7.2\times 10^{-6}$\\
200 &1.82   & 8.91  & $2.1\times 10^{-6}$\\
225 & 1.82 & 8.91 & $3.5\times 10^{-7}$\\
250 & 1.82 & 8.91 & $5.8\times 10^{-7}$\\
\end{tabular} 
\label{tablefour}
\end{table}

In Table \ref{tablefour} we present results for $D$ versus $\Gamma$ for a single imperfect crystal initial condition.   Here $D$ was calculated at $\Gamma=175$.  Next the velocities of the final $\Gamma=175$ configuration were rescaled to $\Gamma=185$ and the system was equilibrate at $\Gamma=185$ and $D$ determined.  This process was repeated for larger $\Gamma$.  We see that $D$ decreases with increasing $\Gamma$ far more slowly than does $D$ for a bcc lattice.  This suggests that $D$ is dominated by the motion of defects and that these defects continue to move even at low temperatures where the ion hopping shown in Fig. \ref{Fig3} is very unlikely.  Note that we expect some variation in these results for $D$ depending on the number and kind of defects present in the initial conditions.

\subsection{Amorphous initial conditions}
\label{subsec.amorphous}

In this subsection, we present results for amorphous initial conditions.   The imperfect crystal initial conditions in Subsection \ref{subsec.frozen} involved a small amount of supercooling, until a configuration froze.  We now consider much greater supercooling.  We start with a liquid configuration of N=8192 ions, that is equilibrated at $\Gamma=175$.  The screening length is $\lambda=1.82a$, $R_{cut}=8.91\lambda$, and the time step is $\Delta t\omega_p=0.12$.  We quench the system instantaneously to a large $\Gamma$ value by rescaling the velocities, and then we evolve the resulting amorphous system at (approximately) constant temperature until the system largely crystalizes.   Note that quenched initial configurations for different $\Gamma$ values were prepared by rescaling the velocities of the same $\Gamma=175$ liquid configuration.  Table \ref{tablefive} lists the time needed to crystalize for different $\Gamma$ values.    This time increases with increasing $\Gamma$ (amount of supercooling).  However, this time only increases approximately linearly with $\Gamma$ for $\Gamma<1500$.  This suggests that diffusion is relatively fast in the amorphous system, and that the amorphous to crystal transition does not involve a large energy barrier.   We find that the system is able to crystalize, even at $\Gamma=1500$ where the temperature is 8.6 times lower than the melting temperature.   However a final run that was quenched to $\Gamma=1750$ was not observed to crystalize before a time $25,000,000/\omega_p$.  We refer to these quenched systems as amorphous.  However, it may be more appropriate to call them polycrystalline because they are observed to have many small crystal domains of different orientation.  These polycrystalline states are observed to undergo rapid transitions to single crystals except at the largest $\Gamma$, see below.    

\begin{table}
\caption{Approximate time $t_0$ for amorphous systems to crystalize, after the systems have been instantaneously quenched from $\Gamma=175$ to different $\Gamma$ values, see text. The number of ions is $N$.}
\begin{tabular}{lll}
$N$& $\Gamma$ & $t_0\omega_p$\\
 \hline
8192 & 500 & 24,000\\
8192 & 600 &  47,000\\
8192 & 700 & 142,000\\
8192 & 1000 & 240,000\\
8192 & 1500 & 390,000\\
8192 & 1750 & $>25,000,000$\\
27648 & 500 & 400,000\\
27648 & 1000 & $>6,000,000$\\
\end{tabular} 
\label{tablefive}
\end{table}

\begin{figure}[ht]
\begin{center}
\vskip.1in
\includegraphics[width=3.25in,angle=0,clip=true] {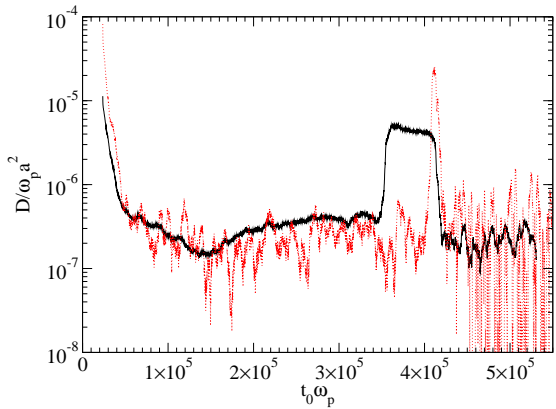}
\caption{(Color on line) Diffusion constant $D$ versus simulation time $t_0$, see Eq. \ref{D'}, for an amorphous system of $N=27648$ ions at $\Gamma=500$.  The diffusion constant $D$ is calculated using a time difference of $t=59,000/\omega_p$ (solid black line) and $t=4720/\omega_p$ (dotted red line) in Eq. \ref{D'}. The sample was prepared by instantaneously quenching a liquid from $\Gamma=175$ to $\Gamma=500$.}
\label{Fig6}
\end{center}
\end{figure}

To study finite size effects we now consider larger systems with $N=27648$ ions.  We start with a liquid configuration equilibrated at $\Gamma=175$, and quench the system instantaneously to $\Gamma=500$.  Figure \ref{Fig6} shows $D$ versus simulation time $t_0$.  The diffusion constant $D$ first decreases rapidly with time as the quenched system anneals.  Next $D$ remains more or less constant for a long time.  Suddenly near $t_0=400,000/\omega_p$ there is a large spike in $D$.  We calculate $D$ with both $t=59000/\omega_p$ and $4720/\omega_p$.  The larger $t$ gives $D$ with less statistical noise, while the smaller $t$ gives better time resolution and shows that the event near $t_0=400,000/\omega_p$ is very rapid.

\begin{figure}[ht]
\begin{center}
\vskip.1in
\includegraphics[width=3.25in,angle=0,clip=true] {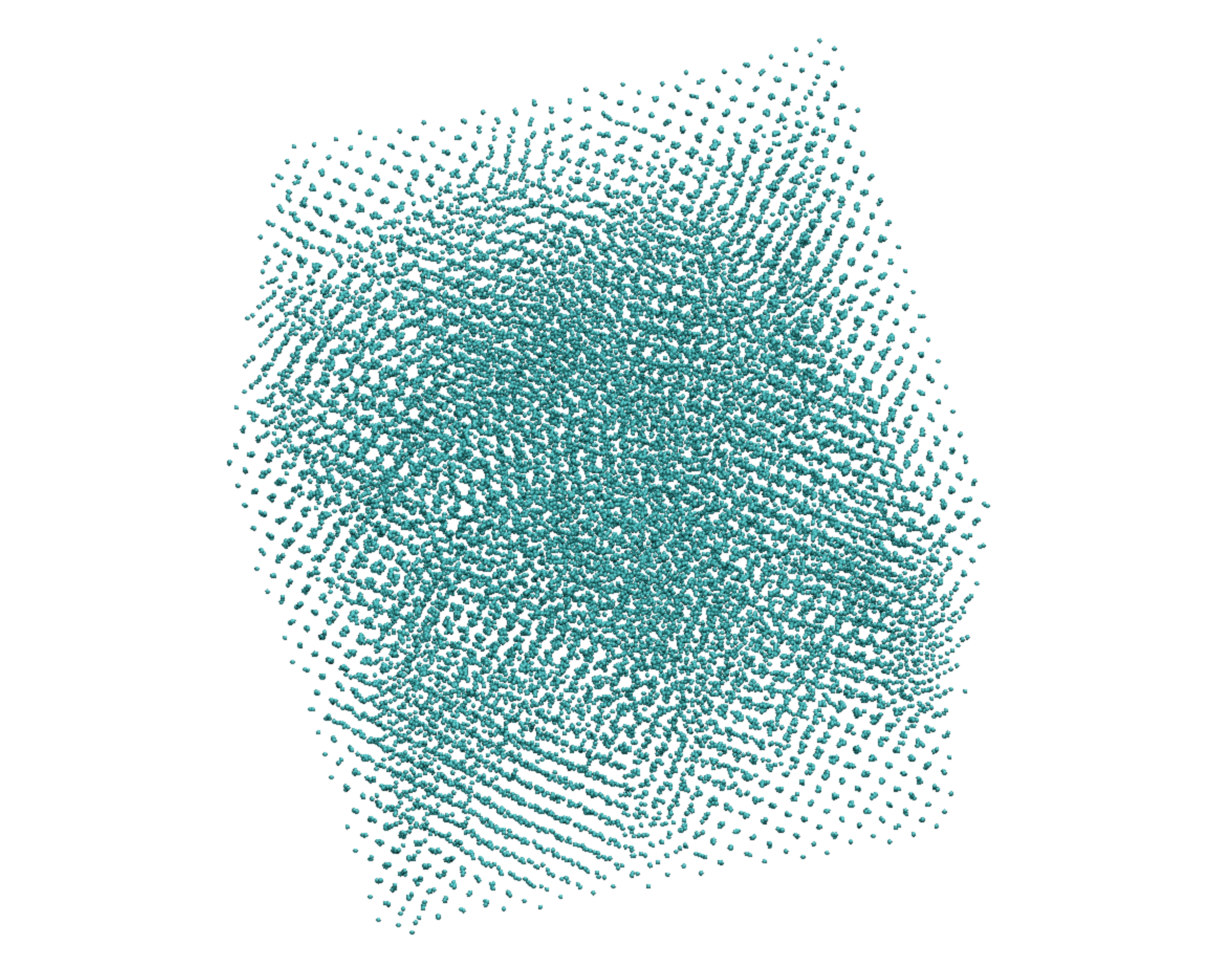}
\caption{(Color on line) Configuration of 27648 ions at $\Gamma=500$ after a simulation time $t_0=350,000/\omega_p$.  The sample was prepared by instantaneously quenching a liquid from $\Gamma=175$ to $\Gamma=500$.   Figure plotted using VMD \cite{VMD}. }
\label{Fig7}
\end{center}
\end{figure}

\begin{figure}[ht]
\begin{center}
\vskip.1in
\includegraphics[width=3.25in,angle=0,clip=true] {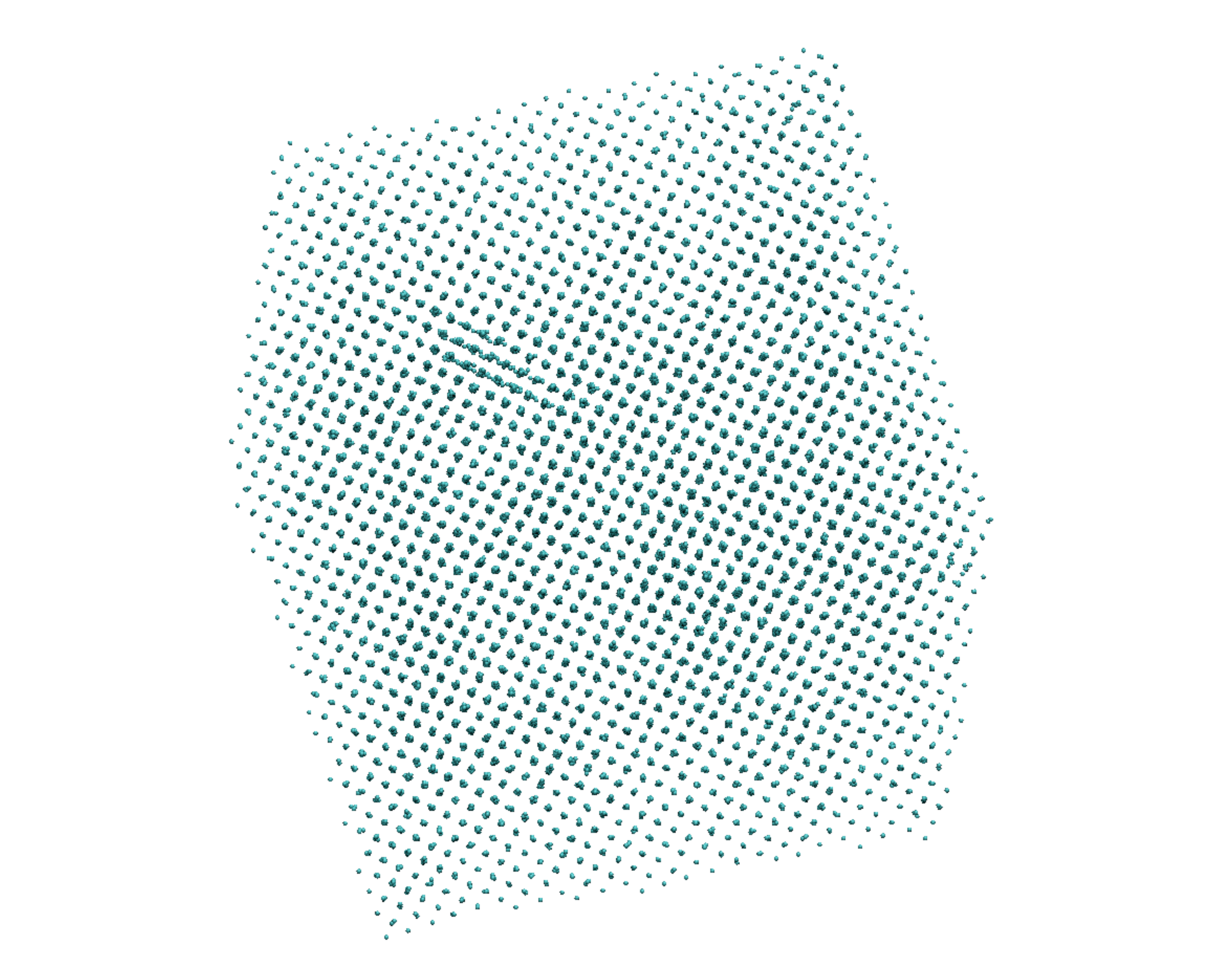}
\caption{(Color on line) Configuration of 27648 ions at $\Gamma=500$ after a simulation time $t_0=450,000/\omega_p$.  The sample was prepared by instantaneously quenching a liquid from $\Gamma=175$ to $\Gamma=500$.   Figure plotted using VMD \cite{VMD}. }
\label{Fig8}
\end{center}
\end{figure}

\begin{figure}[ht]
\begin{center}
\vskip.1in
\includegraphics[width=3.6in,angle=0,clip=true] {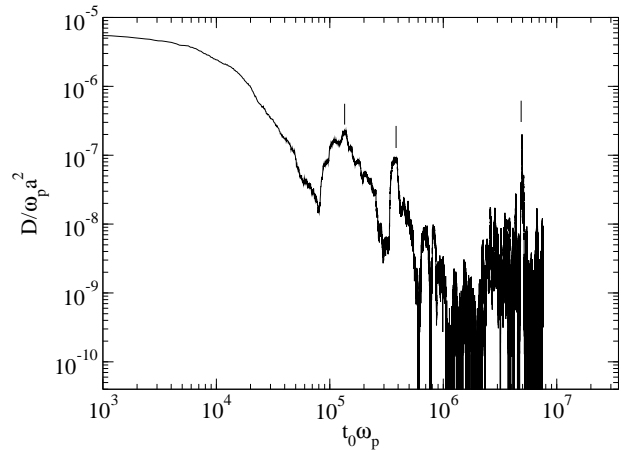}
\caption{(Color on line) Diffusion constant $D$ versus simulation time $t_0$ for a 27648 ion system  at $\Gamma=1000$, using $t=59000/\omega_p$.   The vertical lines mark diffusion features that are also indicated in Fig. \ref{Fig10}.  The system was prepared by instantaneously quenching a liquid from $\Gamma=175$ to $\Gamma=1000$. }
\label{Fig9}
\end{center}
\end{figure}

\begin{figure}[ht]
\begin{center}
\vskip.1in
\includegraphics[width=3.6in,angle=0,clip=true] {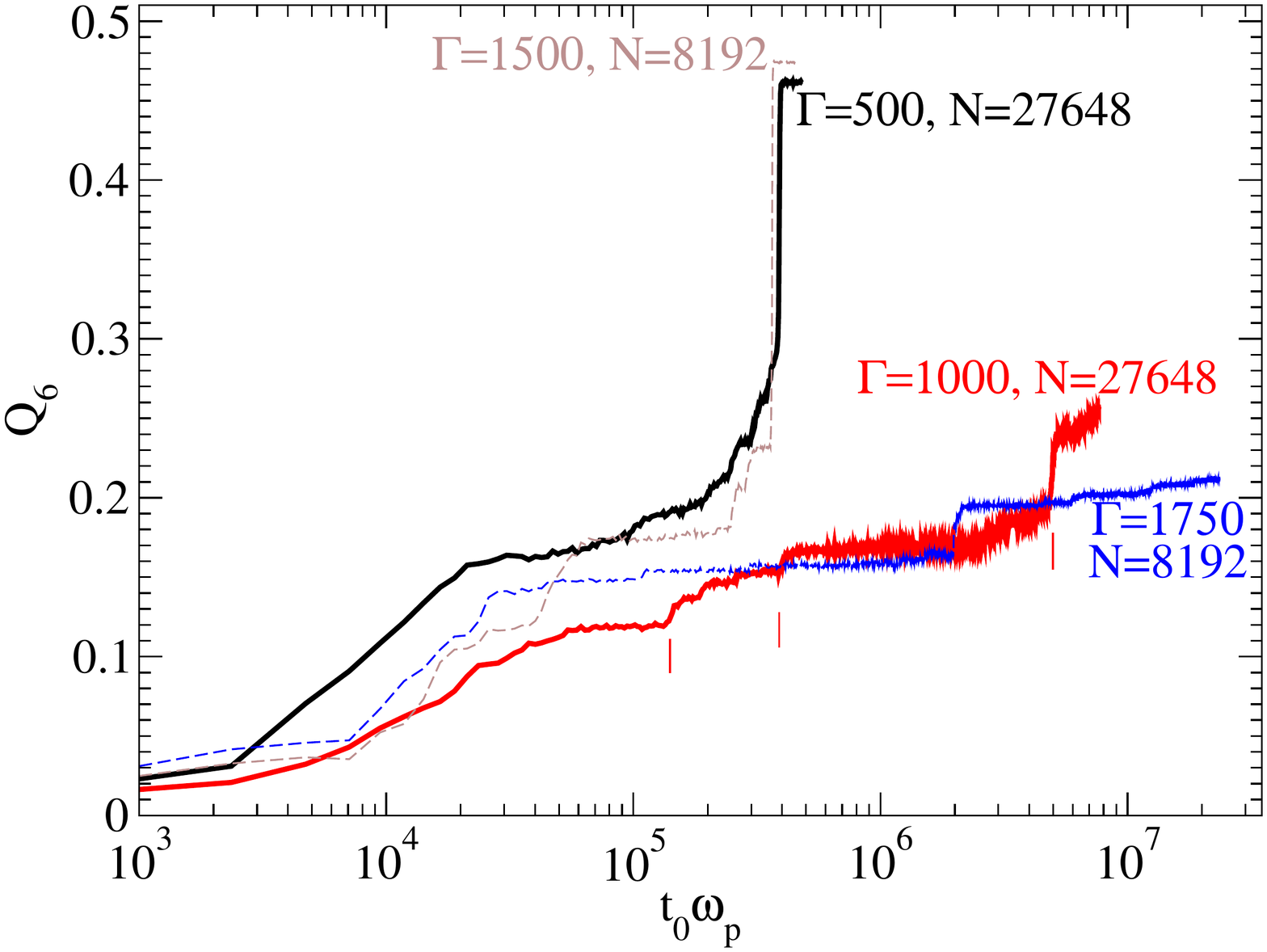}
\caption{(Color on line) Bond angle metric $Q_6$, see Eqs. \ref{Q_lm},\ref{Qbar},\ref{Q_l}, versus simulation time $t_0$ for amorphous systems that were instantaneously quenched from a $\Gamma=175$ liquid at $t_0=0$.  The number of ions in the simulation $N$ and coulomb parameter $\Gamma$ are indicated.  The vertical red lines, for $\Gamma=1000$ and $N=27648$, indicate diffusion features that are seen in Fig. \ref{Fig9}.}
\label{Fig10}
\end{center}
\end{figure}

The configuration of the system just before the event is shown in Fig. \ref{Fig7}.  The system is seen to be in a polycrystalline state with many small crystal domains.  Figure \ref{Fig8} shows the configuration of the system just after the event.  Now the system is an imperfect single crystal.  This demonstrates that diffusion is fast enough, at least at $\Gamma=500$, for the system to crystalize.  Finally in Fig. \ref{Fig9}, we show $D$ as a function of simulation time $t_0$ for an $N=27648$ ion system quenched to $\Gamma=1000$.  Again $D$ starts off large and decreases rapidly as the system starts to equilibrate.  Three large peaks are observed in $D$ near $t_0=150,000$, $400,000$, and $5\times 10^6/\omega_p$.  These correspond to events where small micro-crystals rearrange and grow and the bond angle metric $Q_6$ increases, as we discuss below, see Fig. \ref{Fig10}.  However the system is still polycrystalline after the events. 

To quantify the crystalline order in these simulations, we consider a metric based on bond angles \cite{steinhardt,steinhardt2,bondangles}, see also \cite{klumov}.  Ion $i$ is said to be bonded to ion $j$ if it is within a distance $b=2.44a$ that corresponds to a minimum in the radial distribution function $g(r)$.  This distance is chosen to include the eight nearest neighbors and six next nearest neighbors in a perfect body centered cubic lattice.  Let $\theta_{ij}$ and $\phi_{ij}$ be the polar and azimuthal angles of the radius from ion $i$ to ion $j$.  We calculate the spherical harmonic,
\begin{equation}
Q_{lm}(\hat r_{ij}) = Y_{lm}(\theta_{ij},\phi_{ij}),
\label{Q_lm}
\end{equation}
and average over all $\approx 14 N$ bounds for a given configuration,
\begin{equation}
\bar Q_{lm} =\langle Q_{lm}(\hat r_{ij})\rangle\, .
\label{Qbar}
\end{equation}
This quantity depends on the orientation of a crystal lattice with respect to the simulation volume.  Therefore, we calculate the rotationally invariant quantity $Q_l$ \cite{steinhardt,bondangles},
\begin{equation}
Q_l=\Bigl[\frac{4\pi}{2l+1}\sum_{m=-l}^l|\bar Q_{lm} |^2\Bigr]^{1/2}\, .
\label{Q_l}
\end{equation}
This provides a measure of the crystalline order of a configuration.  In general, $Q_l$ is small for a liquid or amorphous configuration and $Q_l$ is large for a perfect crystal.  Our calculations of $Q_l$, for a range of even $l$, show that $Q_6$ is most sensitive to crystalline order.  We find that
\begin{equation}
Q_6 = 0.51069
\end{equation}
for a perfect bcc lattice and $Q_6=0.57452$ for a perfect face centered cubic lattice, see also \cite{bondangles}.  Note that $Q_l$ is small for odd $l$.  In Fig. \ref{Fig10} we show $Q_6$ versus simulation time $t_0$.  In general, $Q_6$ grows with $t_0$.  However, the amount of time necessary for $Q_6$ to grow can increase strongly with increasing system size $N$ or $\Gamma$.  A plateau near $Q_6\approx 0.17$ is seen for all four systems in Fig. \ref{Fig10}.  This suggests a possible metastable intermediate state.  The simulations with $N=27648$ at $\Gamma=500$ and $N=8192$ at $\Gamma=1500$ show a rapid rise in $Q_6$, near $t_0=4\times 10^5 /\omega_p$, during transitions to single crystals.  For $N=27648$ at   $\Gamma=1000$ and $N=8192$ at $\Gamma=1750$, $Q_6$ is increasing with time.   These systems have not yet evolved to single crystals.  However, the continued rise of $Q_6$ with time strongly suggests that these systems will evolve to single crystals at later times.  In summary, the continued rise of $Q_6$ with time, as shown in Fig. \ref{Fig10}, demonstrates that these quenched systems are evolving with time towards single crystals, and that they are unlikely to remain amorphous. 

We conclude that an amorphous solid will not form even with large amounts of super cooling, where the temperature is rapidly quenched by up to a factor of 10 below the melting temperature.  Instead, diffusion is fast enough so that the system will form a regular crystal.   Our results strongly suggest that Coulomb solids in the interior of cold white dwarf stars and the crust of neutron stars will be crystalline and not amorphous.  This is consistent with observations of rapid crust cooling of neutron stars following extended periods of accretion \cite{NScrustcooling,crustcooling2,crustcooling3,crustcooling4}.  This rapid cooling implies a high crust thermal conductivity, that agrees with the conductivity of a regular crystal, and is larger than the conductivity expected for an amorphous solid.

\section{Conclusions}
\label{sec.conclusion}
Diffusion in coulomb crystals can be important for the structure of the crust of neutron stars.  In this paper, we perform molecular dynamics simulations of one component coulomb crystals to study the diffusion constant $D$.  We find that $D$ is non-zero, at least near the melting temperature, and that $D$ for Coulomb crystals with relatively soft-core $1/r$ interactions is in general larger than $D$ for Lennard-Jones or other solids with harder-core (more singular) interactions.  

We find that diffusion, for simulations that start from a perfect body-centered-cubic lattice, involves the exchange of ions in ring-like configurations.  Here ions ``hop" in unison with one ion replacing another without the formation of long lived vacancies.  This may be true because vacancy formation is strongly suppressed     because of the large pressure.  The diffusion constant $D$ decreases rapidly, for temperatures below the melting point, suggesting that these ring-like configurations have a high activation energy.

We also calculate diffusion for simulations that start from imperfect crystal initial conditions.  Here a liquid configuration, at a temperature somewhat below the melting point, spontaneously freezes to an (in general) imperfect crystal that may contain defects such as dislocations and grain boundaries.  Note that these configurations involve one or more micro-crystals and are not amorphous.  For these systems, $D$ is larger than $D$ for perfect bcc lattice configurations and decreases more slowly with decreasing temperature.  This suggests that $D$ for imperfect crystals is dominated by the motion of the crystal defects rather than the hopping of ions in a perfect crystal.  Therefore, observations of $D$ may help characterize the imperfections in a Coulomb crystal.  

Finally, we studied diffusion in ``amorphous'' systems where the temperature was instantaneously quenched to much lower values.  We find that $D$ is large.  Indeed most of our amorphous simulations are observed to spontaneously transform to either a single crystal or a small number of crystal domains.  This strongly suggests that Coulomb solids in white dwarf and neutron stars are crystalline, rather than amorphous.  This is in agreement with X-ray observations of rapid neutron star crust cooling that imply a large thermal conductivity. 

It is an important open problem to determine the equilibrium distribution of defects in a Coulomb crystal.  It may be difficult to determine this directly from molecular dynamics simulations because it can take a very long time for defects to equilibrate.  However, we find that diffusion is relatively fast.  This suggests that astrophysical coulomb solids will have had plenty of time to anneal to nearly perfect crystals with relatively few defects.   Finally, the diffusion constant that we find for a pure bcc lattice may provide a lower limit on $D$ for an equilibrated system.  This is because the presence of defects is only expected to increase $D$ over that for a perfect crystal.  In future work we plan to study $D$ for multicomponent Coulomb solids.  For a given species $i$, we expect a rich behavior for the diffusion constant $D_i$ depending on how the charge of an ion $Z_i$ compares to the average charge of the ions that make up the crystal lattice.

We thank Andrey Chugunov for very helpful comments.  This research was supported in part by DOE grant DE-FG02-87ER40365 and by the National Science Foundation through TeraGrid resources provided by National Institute for Computational Sciences under grant TG-AST100014.

\vfill\eject

\end{document}